\documentclass[prd,twocolumn,tightenlines,superscriptaddress,nofootinbib]{revtex4}
\usepackage{amssymb,latexsym}
\usepackage{amsmath,amsbsy,bbm}
\usepackage{epsfig,bm}
\usepackage{graphicx,comment}
\usepackage{color}
\usepackage[normalem]{ulem}
\begin{document}

\title{Investigation of a universal behavior between N\'eel temperature 
and staggered magnetization density for a three-dimensional quantum 
antiferromagnet}

\author{M.-T. Kao}
\author{F.-J. Jiang}
\email[]{fjjiang@ntnu.edu.tw}
\affiliation{Department of Physics, National Taiwan Normal University, 
88, Sec.4, Ting-Chou Rd., Taipei 116, Taiwan}

\vspace{-2cm}

\begin{abstract}
We simulate the three-dimensional quantum Heisenberg model with a spatially
anisotropic ladder pattern using the first principles Monte Carlo method. Our
motivation is to investigate quantitatively the newly established 
universal relation $T_N/\sqrt{c^3}$ $\propto$ ${\cal M}_s$ near the 
quantum critical point (QCP) associated with dimerization. Here $T_N$, 
$c$, and ${\cal M}_s$ are the N\'eel temperature, the spinwave velocity, 
and the staggered magnetization density, respectively. For all the 
physical quantities considered here, such as $T_N$ and ${\cal M}_s$, 
our Monte Carlo results agree nicely with the corresponding results 
determined by the series expansion method. In addition, we find it is likely
that the effect of a logarithmic correction, which should be present in 
(3+1)-dimensions, to the relation $T_N/\sqrt{c^3}$ $\propto$ ${\cal M}_s$ 
near the investigated QCP only sets in significantly in the region with 
strong spatial anisotropy. 
\end{abstract}

%\pacs{12.39.Fe, 75.10.Jm, 75.40.Mg, 75.50.Ee}

\maketitle

{\bf Introduction}.---
While being the simplest models, Heisenberg-type models provide 
qualitatively, or even quantitatively useful information 
regarding the properties of cuprate materials. For example, 
the spatially anisotropic quantum Heisenberg model with different 
antiferromagnetic couplings in the 1 and 2 directions is demonstrated to 
be relevant for the underdoped cuprate superconductor
YBa$_2$Cu$_3$O$_{6.45}$ \cite{Hinkov2007,Hinkov2008}. Specifically, it 
is argued that this model provides a possible mechanism for the newly 
discovered pinning effects of the electronic liquid crystal in 
YBa$_2$Cu$_3$O$_{6.45}$ \cite{Pardini08}. Because of their phenomenological 
importance, these models continue to attract a lot of attention 
analytically and numerically. In addition to being relevant to 
real materials, Heisenberg-type models on geometrically nonfrustrated
lattices are important from a theoretical point of view as well. This is 
because these models can be simulated very efficiently using first principles 
Monte Carlo methods. Hence they are very useful in exploring ideas and 
examining theoretical predictions 
\cite{Sac00,Voj00,Sac01,Tro02,Matsumoto02,Hog03,Wan05,Ng06,Alb08}.

Recently a new universal behavior between the thermal and quantum properties
of (3+1)-dimensional dimerized quantum antiferromagnets has been
established \cite{Oti12, Kul11}. Specifically, using the relevant field 
theory, it is shown that the N\'eel temperature $T_N$ can be related to the 
staggered magnetization density ${\cal M}_s$ near a quantum critical 
point (QCP). This new universal property is then compared with experimental 
data for TlCuCl$_3$ in Ref.~\cite{Rue08} and the agreement is impressive. 
In addition, 
in Ref.~\cite{Oti12} the relevant series expansion calculations are performed 
for the (3+1)-dimensional ladder-dimer quantum antiferromagnet. The 
obtained results match reasonably well with the corresponding field theory 
predictions. Similar behavior was obtained in Monte Carlo simulations of 
\cite{Jin12} with various kinds of model.    

Motivated by this newly established universal relation between thermal and 
quantum properties close to a QCP as well as to study this scaling behavior 
quantitatively, we simulate the (3+1)-dimensional ladder-dimer quantum 
Heisenberg model using the first principles Monte Carlo method. The relevant 
quantities such as $T_N$, ${\cal M}_s$, and the spinwave velocity $c$ are 
determined with high precision. We find that our results agree nicely with 
the series expansion calculations presented in Ref.~\cite{Oti12}. 
In particular, with an empirical fitting ansatz, our Monte Carlo data imply 
that the effect of a logarithmic correction, which should be present in 
(3+1)-dimensions, to the relation $T_N/\sqrt{c^3}$ $\propto$ ${\cal M}_s$ near 
the considered QCP only sets in significantly in the region with strong 
spatial anisotropy.

\begin{figure}
\begin{center}
\includegraphics[width=0.25\textwidth]{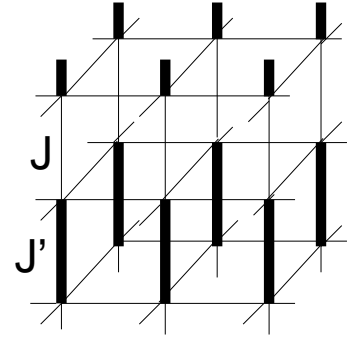}
\end{center}
\caption{The (3+1)-dimensional spatially anisotropic quantum Heisenberg model considered in
this study.}
\label{fig1}
\end{figure}

{\bf Microscopic Model and Corresponding Observables}.---
The three-dimensional quantum Heisenberg
model considered in this study is defined by the Hamilton operator
\begin{eqnarray}
\label{hamilton}
H = \sum_{\langle xy \rangle}J\,\vec S_x \cdot \vec S_{y}
+\sum_{\langle x'y' \rangle}J'\,\vec S_{x'} \cdot \vec S_{y'},
\end{eqnarray}
where $J$ ($J'$) is the antiferromagnetic exchange coupling connecting
nearest neighbor spins $\langle  xy \rangle$ ($\langle x'y' \rangle$). 
The model described by Eq.~(1) and studied here is illustrated in fig. 1. 
To investigate the newly established universal behavior
between $T_N$ and ${\cal M}_s$ near the critical point induced by 
dimerization, the spin stiffnesses in all spatial directions, which 
are defined by
\vskip-0.5cm
\begin{eqnarray}
\rho_{si} = \frac{1}{\beta L_1L_2L_3}\langle W^2_{i}\rangle
,\end{eqnarray}
are measured in our simulations. Here $\beta$ is the inverse temperature, 
$L_i$ refers to the spatial box size in the $i$ direction, 
and $\langle W^2_{i} \rangle$ with $i \in \{1,2,3\}$ is the winding number 
squared in the $i$ direction. In addition, the observable 
$\langle (m_s^z)^2 \rangle$ is recorded in our calculations as well in 
order to determine ${\cal M}_s$. Here $m_s^z$ is the $z$ component of the 
staggered magnetization 
$\vec{m}_s = \frac{1}{L_1L_2L_3}\sum_{x}(-1)^{x_1 + x_2 + x_3}\vec{S}_x$.
To perform the investigation, using the stochastic series expansion
algorithm (SSE) with operator-loop update \cite{San99}, we have carried out large scale 
Monte Carlo simulations with various inverse temperatures and box 
sizes $L$ at several values of $J'/J$ (We use $L_1$ = $L_2$ = $L_3$ in most 
of our simulations and $J$ is set to be 1.0 throughout the calculations). 
Notice that, since the established QCP induced by 
dimerization is at $(J'/J)_c \sim 4.0$ \cite{Noh05}, we have performed our 
calculations for $2.5 \le J'/J \le 4.0$. First of all, let us focus on our 
results of determining $T_N$.

{\bf Determination of the N\'eel Temperatures}.---
To calculate the N\'eel temperatures $T_N$ for which the long-range
antiferromagnetic order is destroyed for $T > T_N$, at each fixed 
$J'/J$ = 2.5, 3.0, 3.25, 3.375, 3.5, 3.625, 3.75, and 3.875, we have performed simulations 
by varying $T$ for $L$ = 8, 12, 16,..., 36, 40. Further, the numerical values 
of $T_N$ are obtained by employing the standard finite-size scaling analysis 
to the relevant observables. Specifically, near $T_N$ and for the 
observables $\rho_{si}L$ with $i\in\{1,2,3\}$, the curves of different $L$ 
as a function of $T$ should tend to intersect at $T_N$. Interestingly, we find 
that at each considered $J'/J$ the correction to scaling for these 
observables is negligible when the relevant data points with $L \ge 20$ are 
employed in the analysis. In other words, our data can be described 
well by the expected leading scaling ansatz. Specifically, the ansatz employed 
in our finite-size scaling analysis is of the form $g(x)$, where $g$ is a
smooth function of the parameter $x$ and $x$ contains a factor linear in 
$(T-T_N)/T_N$.
Indeed, by applying the fourth order Taylor expansion of the expected leading 
scaling ansatz 
to $\rho_sL = (\rho_{s1}+\rho_{s2})L/2$, we arrive at $T_N = 0.7751(2)$ for 
$J'/J = 3.5$ (top panel of fig.~2). Using a third order Taylor expansion of 
the leading scaling form leads to a value of $T_N$ which agrees nicely with 
$T_N = 0.7751(2)$. Employing the same procedure, the value of $T_N$ 
determined from $\rho_{s3}L$ for $J'/J$ = 3.5 is given by $0.7750(2)$ 
(bottom panel of fig.~2). Notice that the $T_N$ obtained from these two 
different
observables agree with each other quantitatively. The $T_N$ at other couplings 
$J'/J$ are calculated with the same strategy and table 1 summarizes our 
results of determining the values of $T_N$ at the considered couplings 
$J'/J$. Notice a bootstrap resampling method is employed in
obtaining the results in table 1. In particular, the quoted errors  
are determined by a conservative estimate based on
the standard deviations of the fits with good quality.
Later these determined $T_N$ will be used in examining the universal 
behavior between $T_N$ and ${\cal M}_s$ near the QCP associated with 
dimerization.

\begin{figure}
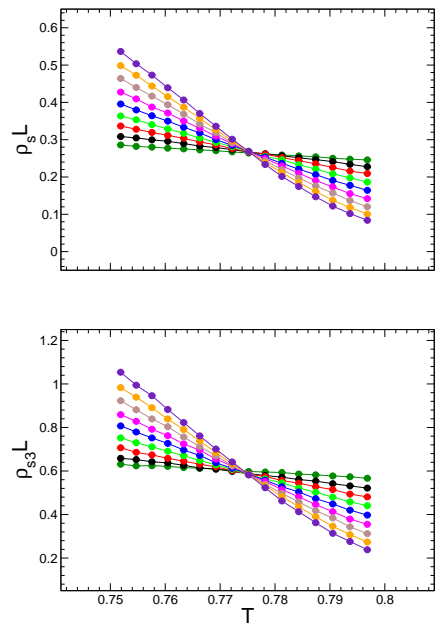

\begin{center}
\vbox{
\includegraphics[width=0.315\textwidth]{rhosL_J3.5.eps}
\vskip0.3cm
\includegraphics[width=0.315\textwidth]{rhos3L_J3.5.eps}
}
\end{center}
\caption{Monte Carlo data of $\rho_{s} L$ (top panel) and
$\rho_{s3} L$ (bottom panel)
for $J'/J=3.5$. The lines are added to guide the eye. }
\label{fig2}
\end{figure}

\begin{table}
\label{tab1}
\begin{center}
\begin{tabular}{ccccc}
\hline
{\text{observable}} & $J'/J $ & $T_N$ & $J'/J$& $T_N$\\
\hline
\hline
$\rho_{s}L$ &  2.5 & 1.0014(2) & 3.5 & 0.7751(2)\\
\hline
$\rho_{s3}L$  & 2.5 & 1.0014(2) &  3.5 & 0.7750(2)\\
\hline
$\rho_{s}L$  &  3.0 & 0.9317(2) & 3.625 & 0.7087(3)\\
\hline
$\rho_{s3}L$  & 3.0 & 0.9316(2) & 3.625 & 0.7086(3)  \\
\hline
$\rho_{s}L$  & 3.25  & 0.8690(2) & 3.75  & 0.6197(2) \\
\hline
$\rho_{s3}L$  & 3.25 & 0.8689(2) & 3.75 & 0.6193(3)\\
\hline
$\rho_{s}L$  & 3.375  & 0.8270(2) & 3.875  & 0.4853(3)\\
\hline
$\rho_{s3}L$  & 3.375 & 0.8269(2) & 3.875 & 0.4849(4) \\
\hline
\hline
\end{tabular}
\end{center}
\caption{The numerical values of $T_N$ at $J'/J$ = 2.5, 3.0, 3.25, 3,375, 
3.5, 3.625, 3.75, 3.875 determined by applying the leading scaling ansatz 
to the relevant observables. All the $\chi^2/{\text{DOF}}$ of these fits are 
smaller than 1.6.  }
\end{table}

\begin{figure}[ht!]
\begin{center}
\includegraphics[width=0.33\textwidth]{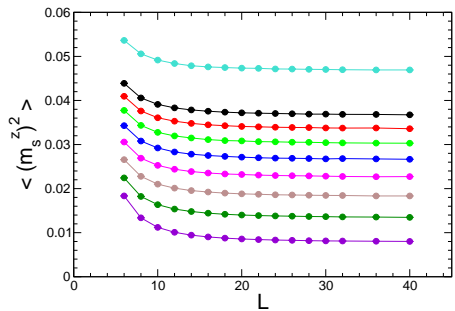}
\end{center}
\caption{Monte Carlo data of $\langle (m_s^z)^2 \rangle$ as functions of $L$. 
From top to bottom the corresponding values of $J'/J$ for these curves are 
2.5, 3.0, 3.125, 3.25,..., 3.75, and 3.875, respectively. The lines are added 
to guide the eye. }
\label{fig4}
\end{figure}

{\bf Determination of the staggered magnetization density}.--- 
To calculate ${\cal M}_s$, we have measured the observable 
$\langle (m_s^z)^2 \rangle$. Specifically, by extrapolating the 
zero-temperature $\langle (m_s^z)^2 \rangle$ at finite lattice size to the 
bulk value $(m_s^z)^2(\infty)$, ${\cal M}_s$ can then be obtained from 
${\cal M}_s$ = $\sqrt{3 (m_s^z)^2(\infty)}$. Notice that to determine ${\cal M}_s$ 
by this method one needs the zero-temperature values of 
$\langle (m_s^z)^2 \rangle$. We have carried out trial runs for $L=20$ with 
$\beta J$ = 20 and $\beta J$ = 40 at $J'/J$ = 2.5, 3.0, 3.125, 3.25, 3.375, 
3.5, 3.625, 3.75, 3.875. The obtained values of $\langle (m_s^z)^2 \rangle$ for 
these two different inverse temperatures $\beta$ at all the considered
couplings $J'/J$ agree reasonably well. Hence the extrapolation using the 
data of $\langle (m_s^z)^2 \rangle$ calculated with $\beta J = L$ in the 
simulations should lead to correct results. Indeed it has been demonstrated in 
Ref.~\cite{Jin12} that the extrapolated values of $\langle (m_s^z)^2 \rangle$ 
for various couplings $J'/J$, determined with the data obtained from 
simulations employing $\beta J = L$ and $\beta J = 2L$, are consistent with 
each other. Fig.~3 shows our $\langle (m_s^z)^2 \rangle$ data for 
$L = 6,8,10,...,32,36,40$ at the considered $J'/J$. The extrapolation results 
for these data using the ansatz $a+b/L+c/L^2+d/L^3$ are depicted in fig.~4.
In fig.~4 the solid curve is reproduced from Ref. \cite{Oti12} and is the 
fitting result based on series expansion calculations. The agreement between
our Monte Carlo data and series expansion results of ${\cal M}_s$ is 
remarkable.

\vskip1cm

\begin{figure}[ht!]
\begin{center}
\includegraphics[width=0.33\textwidth]{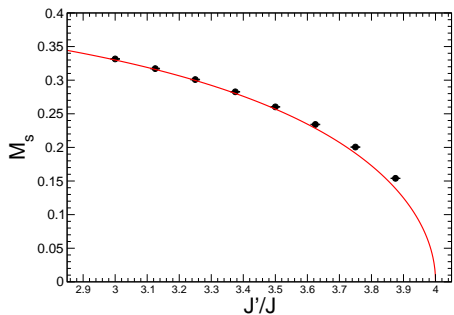}
\end{center}
\caption{Monte Carlo determination of ${\cal M}_s$ as a function of $J'/J$.
The solid curve is reproduced from Ref.~\cite{Oti12} and is the fitting result 
based on series expansion calculations.}
\label{fig5}
\end{figure}

{\bf Determination of the spinwave velocity}.--- 
There are several methods to determine the low-energy constant $c$. Here
we use the idea of winding numbers squared. Specifically, for each $J'/J$ 
we adjust the ratio of $L_1/L_3$ so that all three spatial winding numbers 
squared take approximately the same values. Then we tune $\beta$ in order 
to reach the condition $\langle W_t^2 \rangle \sim \langle W_i^2 \rangle$ 
for $i\in\{1,2,3\}$. Here $ \langle W_t^2 \rangle$ is the temporal winding 
number squared. Once this condition is met, the numerical value of $c$ 
is estimated to be $L/\beta_2 \le c \le L/\beta_1$, where 
$L = (L_1 L_2 L_3)^{1/3}$ and $\beta_1$ ($\beta_2$) stands for the 
largest (smallest) inverse temperature so that the criterion 
$\langle W_{i}^2 \rangle \le \langle W_t^2 \rangle$
($\langle W_{i}^2 \rangle \ge \langle W_t^2 \rangle$) for $i\in \{1,2,3 \}$
is satisfied. For the isotropic case $J'/J = 1.0$, the spinwave theory predicts
$c \sim 1.9091J$ \cite{Oit94}. Remarkably, for a trial simulation with 
$J'/J = 1.0$, 
$L_{1}$ = $L_{2}$ = $L_{3}$ = 20 and $\beta J = 10.476$ 
(hence $L/\beta$ $\sim$ $1.9091 J$),
the ratio of the average of three spatial winding numbers squared and the
temporal winding number squared is 0.994 approximately. This confirms the
validity of calculating $c$ using the idea of winding numbers squared.
For each coupling $J'/J$ studied here, we further consider 
at least two sets of box sizes for which the condition
$\langle W_t^2 \rangle$ $\sim$ $\langle W_i^2 \rangle$ for $i\in\{1,2,3\}$
is satisfied. With this strategy, the numerical values of $c$ obtained
for $J'/J$ = 2.5, 3.0, 3.25, 3.5, 3,375, 3.625, 3.75, 3.875, and 4.0 are 
shown in table~2. The results shown in table~2 imply that the values of 
$c$ at the considered couplings are already convergent to the corresponding
bulk values. Even if some of our determined $c$ have not reached their
bulk values, one expects the deviations to be very small. Hence 
such systematic uncertainty would have little impact on our investigation
of the universal relation between $T_N$ and ${\cal M}_s$.  

\begin{table}
\begin{center}
\begin{tabular}{cccccccc}
\hline
$J'/J $ & $L_1$ & $L_3$ & $c/J$ & $J'/J $ & $L_1$ & $L_3$ & $c/J$ \\
\hline
\hline
2.5 & 22 & 28 & 2.215(8) & 3.5 & 46 & 62 & 2.348(10)\\
\hline
2.5 & 36 & 46 & 2.215(9) &  3.625 & 22 & 30 & 2.360(12)\\
\hline
3.0 & 32 & 42 & 2.282(13) & 3.625 & 34 &46 &2.360(13)\\ 
\hline
3.0 & 44 & 58 & 2.283(11) &  3.75 & 22 & 30 & 2.376(12)\\
\hline
3.25 & 12 & 16 & 2.317(7) & 3.75 & 44 & 60 & 2.378(11)\\
\hline
3.25 & 18 & 24 & 2.317(8) & 3.875 & 16 & 22  & 2.391(7)\\
\hline
3.25  & 24 & 32 & 2.317(11) & 3.875  & 32 & 44 & 2.389(8)\\
\hline
3.375  & 12 & 16 & 2.335(12) & 4.0 & 16 & 22 & 2.408(13)\\
\hline
3.375  & 24 & 32 & 2.334(13) & 4.0 & 32 & 44  & 2.405(15)\\
\hline
3.5 & 34 & 46 & 2.347(12) & 4.0  & 42 & 58 & 2.401(10) \\
\hline
\hline
\end{tabular}
\end{center}
\caption{The numerical values of $c$ obtained through the winding numbers 
squared for various couplings $J'/J$}
\end{table}

~
~
~
~
~

\begin{figure}[ht!]
\begin{center}
\includegraphics[width=0.3\textwidth]{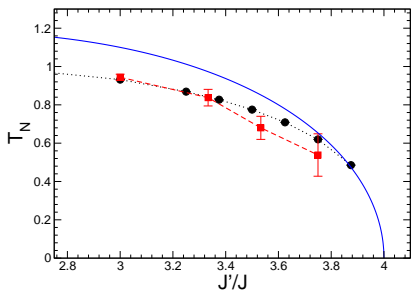}
\end{center}
\caption{Results of $T_N$ as a function of $J'/J$ determined by Monte Carlo 
simulations (solid circles), series expansion (solid squares) as well as 
field theory calculations (solid curve). The field theory and series
expansion results are estimated and reproduced from Ref.~\cite{Oti12}.
The dashed and dotted lines are added to guide the eye. }
\label{fig6}
\end{figure}

{\bf Comparison between theoretical predictions and Monte Carlo results}.---
In Ref.~\cite{Oti12} the following universal relation between 
$T_N$ and ${\cal M}_s$ near a QCP is predicted using the corresponding 
field theory  
\begin{equation}
\label{TNMS}
T_N = \sqrt{\frac{12c^3}{5}}{\cal M}_s
.\end{equation}
Notice the original prediction in Ref.~\cite{Oti12} has $c_1c_2c_3$ instead of
$c^3$ for anisotropic systems. Here $c_i$ refers to the spinwave velocity in
$i$ direction. On the other hand, considering the fact that both $T_N$ and 
${\cal M}_s$ in Eq.~(3) are bulk properties of the system for any given $J'/J$,
it is naturally to use the bulk spinwave velocity $c$ in the prediction. 
The use of $c$ in Eq.~(3) is consistent with the way we determine this 
quantity. As we will demonstrate in the following, Eq.~(3) is valid with our 
interpretation. To verify Eq.~(3), in Ref.~\cite{Oti12} the numerical values of 
$T_N$, $c$, and ${\cal M}_s$ for various couplings $J'/J$ are determined
numerically using the series expansion method. Further, the agreement between
the numerical results from series expansion calculations and the field 
theory prediction is shown to be reasonably good. The quantum Monte Carlo 
determination of $T_N$ and ${\cal M}_s$ for the (3+1)-dimension ladder-dimer 
model is available in Ref.~\cite{Jin12} as well. Notice that to quantitatively 
investigate the relation in Eq.~(3), one needs to additionally calculate $c$. 
This motivates our study presented here. As a first step to quantitatively
study Eq.~(\ref{TNMS}), in fig.~4 we have already compared our Monte Carlo
results for ${\cal M}_s$ with those determined by the series expansion method 
obtained in Ref.~\cite{Oti12}. The consistency of ${\cal M}_s$ calculated with
these two different methods is impressive. Next, we compare our Monte Carlo 
data of $T_N$ with the series expansion results available in 
Ref.~\cite{Oti12}. Such a comparison is 
presented in fig.~5. Near the QCP $J'/J \sim 0.4$, the consistency
between the values of $T_N$ determined by these two different methods is
reasonably good as well. Finally, Eq.~(3) implies that the curve of 
$T_N/\sqrt{c^3}$ as a function of ${\cal M}_s$ should be linear assuming the 
logarithmic correction is not taken into account. In fig.~6 we compute 
$T_N/\sqrt{c^3}$ as a function of ${\cal M}_s$. Indeed qualitatively the curve 
shown in fig.~6 is linear in ${\cal M}_s$. A fit of the $T_N/\sqrt{c^3}$
data for $J'/J = 2.5, 3.0,...,3.875$ in fig.~6 to the expression $a+b{\cal M}_s$ 
leads to $a = 0.0117(20)$ 
which is slightly above the expected value $a = 0$. We attribute such deviation
to the logarithmic correction not taken into account in our analysis.
Since the obtained $a=0.0117(20)$ is only slightly above zero, one expects 
that for the considered parameters $J'/J$, either the effect due to the 
logarithmic correction is small or this correction only sets in significantly 
for the region with much stronger spatial anisotropy. Interestingly, the value
of $b$ obtained from the fit is about half of the predicted value $\sqrt{12/5}$.
This needs further investigation. One possible explanation is that we use
$c^3$ instead of $c_1c_2c_3$ in Eq.~(3).
Without the explicit form of the logarithmic correction, we are not able
to properly describe the data in our analysis. On the other hand,
in the spirit of the expansion in chiral perturbation theory
for Quantum Chromodynamics, it is naturally to include 
${\cal M}_s \log({\cal M}_s)$ as the additional correction. Remarkably, 
we can reach a good result using the ansatz 
$b_1{\cal M}_s + d_1{\cal M}_s\log({\cal M}_s)$ for the fit (dashed line in 
fig.~6). Notice the resulting fitting curves of these two different 
ans\"atze match nicely in the regime where our Monte Carlo data are available.

{\bf Discussions and Conclusions}.---
In this report, we have simulated the three-dimensional ladder-dimer 
quantum Heisenberg model using the first principles Monte Carlo method. Our 
motivation is to investigate quantitatively the newly established universal
relation between $T_N$ and ${\cal M}_s$ near a QCP. We find that for all the 
quantities considered here, such as $T_N$ and ${\cal M}_s$, our Monte Carlo 
calculations agree nicely with the corresponding results determined by the
series expansion method. Assuming Eq.~(3) is correct without considering the 
correction, then $T_N/\sqrt{c^3}$ as a function of ${\cal M}_s$ should vanish 
at ${\cal M}_s$ = 0. We find that the deviation between the extrapolated 
result of $T_N/\sqrt{c^3}$ and zero is of the order $10^{-2}$. This implies 
that either the logarithmic correction is small or this correction only sets 
in significantly for the region with much stronger spatial anisotropy.  
Indeed, our Monte Carlo data of $T_N/\sqrt{c^3}$ can be described well 
by an empirical ansatz $b_1{\cal M}_s + d_1{\cal M}_s\log({\cal M}_s)$. 
Further, the resulting fitting curves of the two different ans\"atze used 
in our analysis match nicely in the regime where our Monte Carlo data are 
available. This confirms that indeed the logarithmic correction only sets 
in significantly for the region beyond what we have studied. Finally,
using the spinwave theory and series expansion results available in 
Refs.~\cite{Oit94,Oit04}, one obtains $T_N \sim 0.944$, 
${\cal M}_s \sim 0.424$, and $c \sim 1.9091 J$ for the isotropic case 
$J'/J =1.0$. The data point of $T_N/\sqrt{c^3}$ and its corresponding
${\cal M}_s$ for $J'/J = 1.0$ is depicted as the square in fig.~6.
It is remarkable that the prediction Eq.~(3) is valid (qualitatively) all 
the way up to ${\cal M}_s \sim 0.4$.
 
~
~
~
~
~

\begin{figure}[ht!]
\begin{center}
\includegraphics[width=0.32\textwidth]{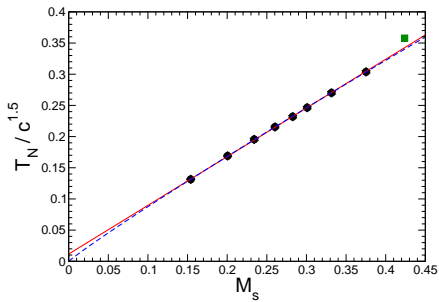}
\end{center}
\caption{Monte Carlo data of $T_N/\sqrt{c^3}$
as functions of ${\cal M}_s$. While the solid line is the result
of fitting the data to the form $a+b{\cal M}_s$, the 
dashed line is obtained using the ansatz 
$b_1{\cal M}_s+d_1{\cal M}_s\log({\cal M}_s)$ for the fit.
The square symbol stands for the result associated
with $J'/J = 1.0$ and is obtained using the calculations 
in Refs.~\cite{Oit94,Oit04}.
}
\label{fig7}
\end{figure}

%\section{Acknowledgments}
%\vskip-0.25cm
Partial support from NSC (Grant No. NSC 99-2112-M003-015-MY3)
and NCTS (North) of R.O.C. is acknowledged. We appreciate greatly
useful discussions with A.~W.~Sandvik and U.-J.~Wiese.

\end{document}